\documentstyle[11pt,newpasp,twoside,epsf]{article}
\def\edcomment#1{\iffalse\marginpar{\raggedright\sl#1\/}\else\relax\fi}
\reversemarginpar

\begin{document}
\title{The Bologna submillisecond pulsar survey}
 \author{Nichi D'Amico}
\affil{Bologna Astronomical Observatory, via Ranzani 1, 40127 Bologna,
   Italy}

\begin{abstract}
Since the discovery of the original millisecond
pulsar, no pulsars with a shorter spin period 
(P$<$1.56 ms) were found.  However, according to the most
popular equations of state, the theoretical limiting spin period of a neutron
star can be much shorter.
On the other hand, most of the large scale searches for millisecond pulsars carried out so far 
were strongly biased
against the detection of ultrashort periodicities.  In this paper we
describe a new large scale pulsar survey with a minimum detectable period much shorter
than previous searches.
\end{abstract}

\vspace*{-0.7truecm}
\section{Introduction}
\vspace*{-0.2truecm}
The discovery of the first millisecond
pulsar (MSP) PSR J1939+2134 (Backer et al, 1982), having a rotational 
period of $\simeq$ 1.56 ms, rised the challenging question of the
limiting spin period of neutron stars (see D'Amico, 1998, and references
therein). A sensitive search for ultrafast pulsars requires 
sophysticated observing equipments, huge data storage devices, and 
supercomputing facilities, so it is not surprising that the minimum 
period observed so far is that of the original brigth millisecond 
pulsar: indeed only periodicities above this value were 
{\it effectively} searched by radioastronomers resulting in the present sample 
of about 80 objects (in the field 
and in Globular Clusters) with periods of the order of few
milliseconds.   The minimum observed period 
is remarkably close to the limiting spin period of a neutron 
star predicted by the so called "stiff" equations of state of the
ultradense matter.  The existence 
of such a pulsar, and its "clock" stability suggests that 
this objects must  be spinning well above the break-up limit of neutron 
stars, implying a much lower degree  of stiffness of the ultradense matter.  
Indeed, other realistic and equally qualified equations of state were 
proposed, resulting in a range of neutron  star limiting spin periods. 
The shortest break-up periods ($\simeq$ 0.6 ms)  are those predicted 
by the so called "soft" equations of state (see Burderi \& D'Amico, 1997, and
reference therein). Possenti et al (1999), showed that Nature could provide
evolutionary paths for spinning a significant amount of neutron stars
up to such extremely high rotational regime.
So, in principle, a search of the submillisecond period range can be used 
to put constraints on the  equation of state of matter at nuclear densities. 
Triggered by these  considerations, we commissioned a new pulsar search 
experiment at the Northern Cross radiotelescope, near Bologna, Italy. 
The experiment has enough time resolution to detect these objects, and 
it is equipped with an online data processing system.  It is
designed such that very narrow pulses as those of the original millisecond 
pulsar can be easily
detected, and it has a similar  sensitivity level for a pulsar spinning near 
the limiting spin period predicted by the softest equation of state 
($\simeq$ 0.6 ms).    Substantially, the present experiment represents 
the first systematic large scale search of the ultrashort period 
range (P$<$1.5 ms).
  
\vspace*{-0.7truecm}
\section{Sensitivity requirements of the experiment}

\begin{figure}
\plotone{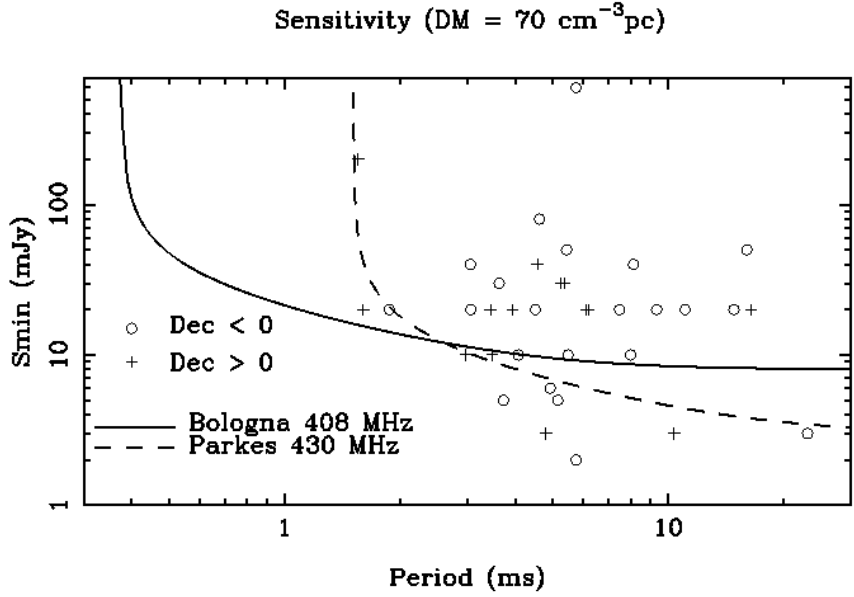}
\caption{Sensitivity profile of the Bologna survey compared with that of the
low frequency Parkes survey (Manchester et al,1996). The positions of the known
millisecond pulsars in the Northern and Southern sky are indicated} 
\end{figure}

The detectability of MSPs is the result of a compromise 
between several requirements. The minimum detectable 
pulsar mean flux density is given by:

\begin{equation}
S_{min} \propto  {k}\frac{S_{SYS}}{\sqrt{\Delta \nu \Delta t}}
{\sqrt{\frac{w_{e}}{P-w_{e}}}} 
\end{equation}

\noindent where S$_{SYS}$ is the system noise equivalent flux density in Jy, $k$ is 
a factor that 
accounts for the adopted detection thereshold and various system losses,
(typically $k$ $\simeq$ 10),  
$\Delta \nu$ is the observed
bandwidth, $\Delta t$ is the integration time, P is the pulsar period, and
$w_{e}$ is the effective pulse
width given by

\begin{equation}
w_{e} = \sqrt{ w^{2} + (\delta t)^{2} + (\frac{DM}{1.2 10^{-4}} 
\frac{ \delta \nu }{ \nu^{3} } )^{2} + (\delta t_{scatt})^2 }
\end{equation}

\noindent where $w$ is the intrinsic pulse width, $\delta t$ is the time resolution
as determined by the sampling time, the post-detection time constant and 
anti-aliasing filter,  $DM$ is the dispersion
measure in cm$^{-3}$pc, $ \delta \nu $ and $\nu$ are respectively the
frequency resolution and the center observing frequency in MHz, and 
$\delta t_{scatt}$ is the broadening of pulses due to multi-path 
scattering in  the interstellar medium.   

The scattering term 
$\delta t_{scatt}$ is roughly proportional to the pulsar distance and 
 scales as $\nu^{-4}$ so, searches at meter wavelengths are usually expected 
to find mainly nearby, high latitudes millisecond pulsars.  The 
instrument used for the present experiment is the E-W arm 
of the Northern Cross radiotelescope, near Bologna, Italy, which
operates at 408 MHz.  The experiment parameters are
summarized in Table 1, and the corresponding sensitivity is shown 
in Fig. 1.

\vskip -0.5truecm
\begin{table}
\begin{center}
\begin{tabular}{|c|c|c|c|c|c|} \hline
\multicolumn{6}{|c|}{Table 1 -- Parameters of the Bologna experiment} \\ \hline
{$\nu$} & {$\Delta\nu$}  & {$\Delta$t} &  {S$_{SYS}$} & {$k$} & {$\delta$t} \\ 
 (MHz)   &   (MHz)        &    (s)      &     (Jy)     &       &    ($\mu$s) \\ \hline
 408     &     4          &    67       &      45      &  10   &     64   \\ \hline
\end{tabular}
\end{center}
\end{table}

\vspace*{-1.1truecm}
\section{Data acquisition and online processing}
\vspace*{-0.2truecm}
In the Northern Cross pulsar system (D'Amico et al, 1996), the IF signal is split 
into a filterbank consisting of 128 channels
each of width 32 kHz.  The outputs are detected, low-pass filtered and
one-bit digitised.   The data acquisition  and the
hardware set-up are controlled by a PC-style microcomputer using a customized 
external bus and a digital I/O expansion unit.  The PC is used to program
the various blocks and log the relevant observation parameters. 
 A special 
purpose card (D'Amico \& Maccaferri, 1994)
is used to manage a data rate up to several megasamples 
per second, and to keep timing precision.   Digital data are
transferred to a realtime data processing subsystem using
a fast link.

The original feature of the realtime data processing subsystem (Fig. 2) is the
dedispersing unit.  This is essentially a large bank of 1-bit addressable
RAM where the raw digital data of each observed beam position are stored.
A special programmable circuitry allows to pick up the the 1-bit samples,
sum them with a given dispersion delay, and output the corresponding
dedispersed time series to a local bus.  Any dedispersed time series 
transferred on the local bus can be read by any of the four CPUs  
available on the same bus and searched for periodicities.  The most
relevant suspect periodicities of each trial DM step are saved into a memory buffer
and finally sorted.   The most significant suspect periodicities detected
on each beam position are saved on a permanent database.  An offline
program is available to selectively search the database, and classify
the suspect by human inspection.
  
\vspace*{-0.3truecm}
\section{Survey status and results}
\vspace*{-0.2truecm}
So far, we observed about 80 \% of the sky region in the declination interval
4$^{o}$ $<$ $\delta$ $<$ 42$^{o}$. We have'nt discovered any pulsar with
a period shorter or similar to that of the original millisecond pulsar.
We have detected 35 known long period pulsars, 5 known
millisecond pulsars and 1 new millisecond pulsar PSR J0030+0451, which was also
discovered at Arecibo (Somer et al, these proceedings). This new millisecond
pulsar has a period P=4.86 ms and a very low DM=4.3 pc$^{-3}$pc.  After the discovery,
made early in 1999, we begun a series of follow-up observations using the Parkes 
radiotelescope. In particular we made long integrations at 70cm in order 
to measure the scintillation parameters and derive the scintillations velocity
(Nigro et al, in preparation).

\begin{figure}
\plotfiddle{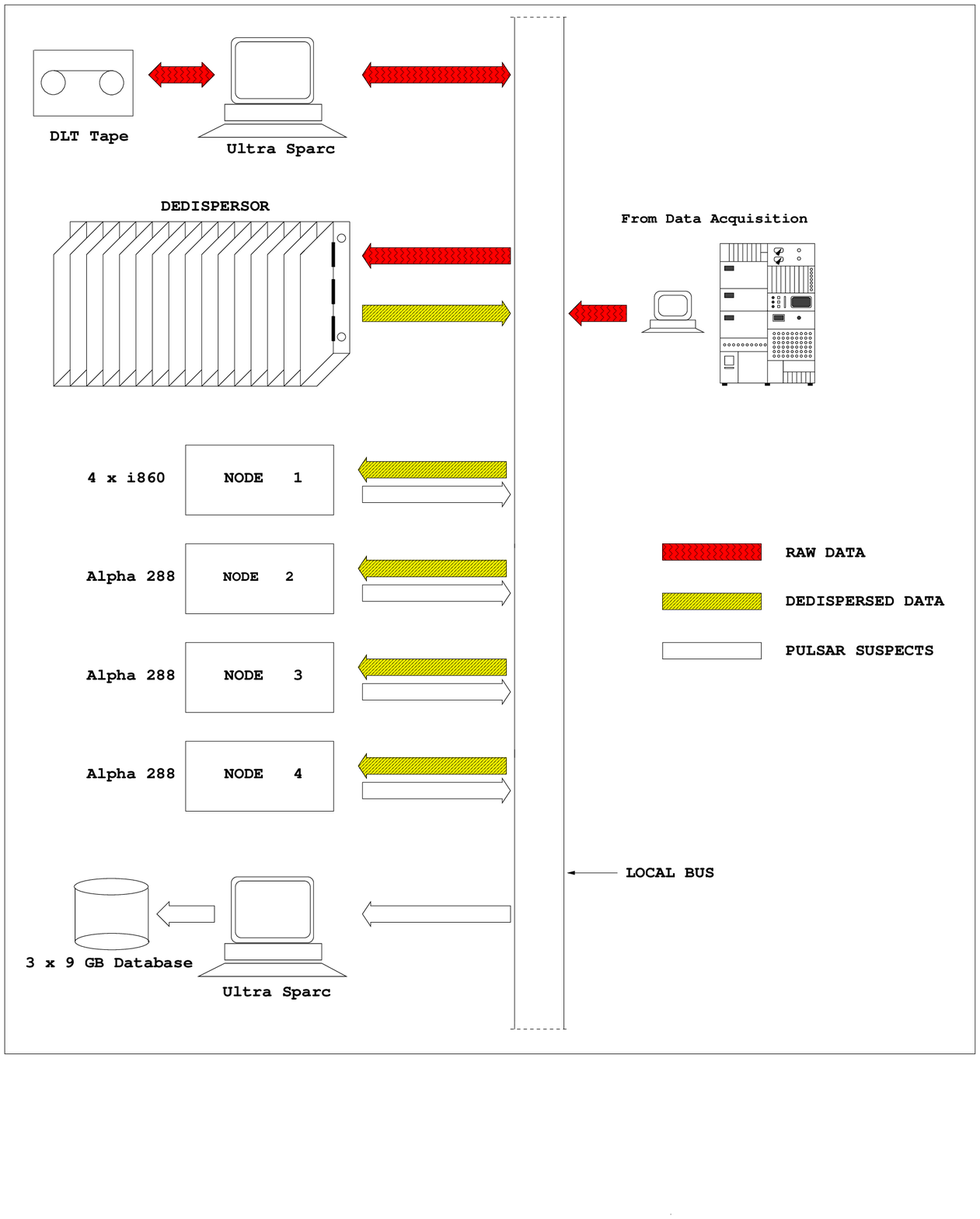}{9.cm}{0}{40}{45}{-110}{-40}
\caption{Block diagram of the data processing subsystem}
\end{figure}

\end{document}